\documentclass{llncs}
\usepackage{makeidx} 
\usepackage{graphicx}
\usepackage{amsmath,amssymb} 
\newcommand{\ket}[1]{ \left |#1\right \rangle}
\makeindex

\begin{document}

\setcounter{page}{1}
\pagestyle{headings}  

\title{\large Unifying Quantum Computation with Projective Measurements only and One-Way Quantum Computation}
\author{Philippe Jorrand, Simon Perdrix}
\institute{Leibniz Laboratory\\46 avenue F\'elix Viallet 38000 Grenoble, France\\\emph{philippe.jorrand@imag.fr, simon.perdrix@imag.fr}}
\maketitle

\begin{abstract}
Quantum measurement is universal for quantum computation (Nielsen \cite{N01}, Raus\-sendorf \cite{R00,R03}). Two models for performing measure\-ment-based quantum computation exist: the one-way quantum computer was introduced by Briegel and Raussendorf \cite{R00}, and quantum computation via projective measurements only by Nielsen \cite{N01}. The more recent development of this second model is based on \emph{state transfers} \cite{P04} instead of teleportation. From this development, a finite but approximate quantum universal family of observables is exhibited, which includes only one two-qubit observable, while others are one-qubit observables \cite{P04}. In this article, an infinite but exact quantum universal family of observables is proposed, including also only one two-qubit observable. 

The rest of the paper is dedicated to compare these two models of measurement-based quantum computation, i.e. one-way quantum computation and quantum computation via projective measurements only. From this comparison, which was initiated by Cirac and Verstraete \cite{VC03}, closer and more natural connections appear between these two models. These close connections lead to a \emph{unified} view of measurement-based quantum computation.

\end{abstract}

\section{Introduction}
Quantum measurement is universal for quantum computation (Nielsen \cite{N01}, Raus\-sendorf \cite{R00,R03}). There exist two models for performing quantum computation with measurements only: one-way quantum computation, introduced by Briegel and Raussendorf \cite{R00}, and quantum computation via projective measurements only, introduced by Nielsen \cite{N01} and improved successively by Leung \cite{L01,L03} and Perdrix \cite{P04}.
One-way quantum computation consists in performing one-qubit measurements on a lattice of qubits initialized in a specific entangled state: the \emph{cluster state}, whereas quantum computation via projective measurements only consists in simulating any unitary transformation using a teleportation-like scheme. 

These two families of measurement-based quantum computations have been recently linked by Cirac and Verstraete \cite{VC03}, who introduced a concept of virtual qubits. We give another approach to closer connections between these two families by considering on the one hand, one-way quantum computation, and on the other hand, quantum computation via measurements only based on state transfer \cite{P04}. 
These connections are established by analyzing how the preparation of the cluster state can be obtained starting from a non-entangled state, while using measurements only. These connections lead to a natural translation of any (one dimensional) one-way quantum computer into a sequence of generalized state transfers and vice versa.

\section{Survey of quantum computation via measurements only based on state transfer}
The computation introduced by Nielsen \cite{N01}, developed by Leung \cite{L01}, is based on teleportation. State transfer is an alternative to teleportation for purpose of computation. State transfer needs less measurements and less auxiliary qubits than teleportation, but in return, state transfer cannot replace teleportation in \emph{non-local} applications. 

\begin{center}
\includegraphics[width=0.5\textwidth]{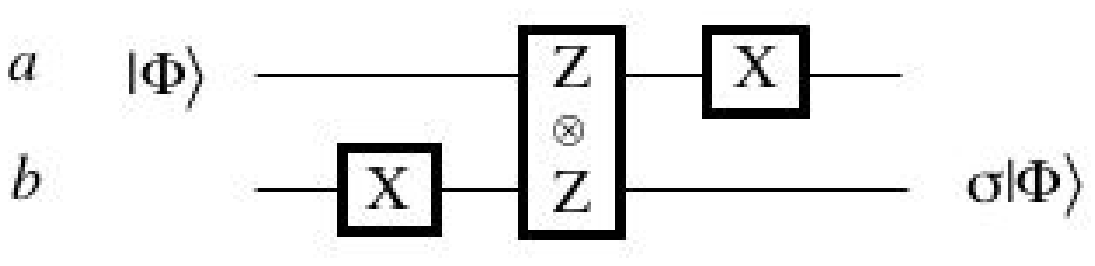}

\emph{\small Figure 1 - State Transfer}
\end{center}

Measurements are defined by the Pauli observables $X$, $Y$ and $Z$.

For a given qubit $a$ and an auxiliary qubit $b$, the sequence of measurements $\{X^{(b)}, Z^{(a)}\otimes Z^{(b)}, X^{(a)}\}$ (see fig. 1), \emph{transfers} the state $\ket{\phi}=\alpha\ket{0}+\beta\ket{1}$ from $a$ to $b$ up to a Pauli operator which depends on the classical outcomes of the measurements.

\begin{center}
\includegraphics[width=0.5\textwidth]{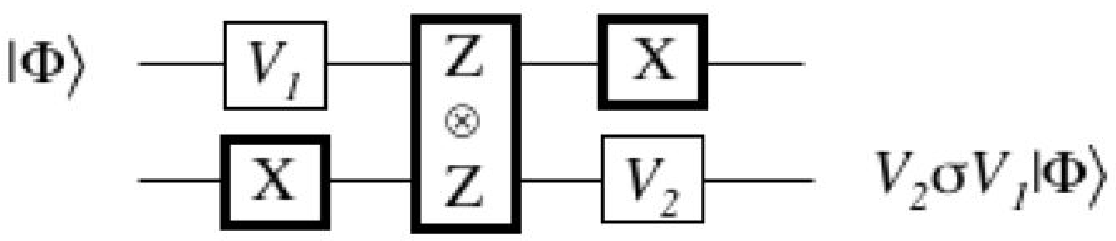}

\emph{\small Figure 2 - State Transfer with additional unitary transformations $V_1$ and $V_2$.}
$$ $$

\includegraphics[width=0.7\textwidth]{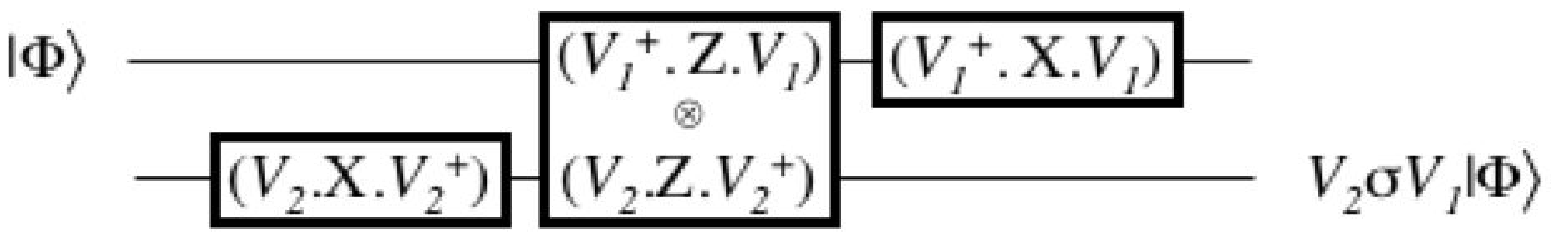}

\emph{\small Figure 3 - Generalized State Transfer.}
\end{center}

By modifying the measurements performed during the state transfer, all $1$-qubit unitary transformations $U$ may be simulated up to a Pauli operator, using generalized state transfers, see fig. 2 and fig. 3 with $V_1=I$ and $V_2=U$. This \emph{step of simulation} of $U$ (i.e. the simulation of $U$ up to a Pauli operator $\sigma$) is followed by a stage of correction which consists in simulating $\sigma$. The reader may reffer to \cite{P04} for details on the stage of correction. 

Generalized state transfers which simulate $H$, $HS^\dag$ and $HT$ are given in fig. 4, 5 and 6, where:
 
\begin{center}
$H=\frac{1}{\sqrt{2}}\left(\begin{array}{cc}
  1 & 1\\
  1 & -1\\
\end{array} \right)$, 
$T=\left(\begin{array}{cc}
  1 & 0\\
  0&e^{\frac{i\pi}{4}}\\
\end{array} \right)$, 
$S=\left(\begin{array}{cc}
  1 & 0\\
  0&i\\
\end{array} \right)$
\end{center}

\begin{center}
\includegraphics[width=0.4\textwidth]{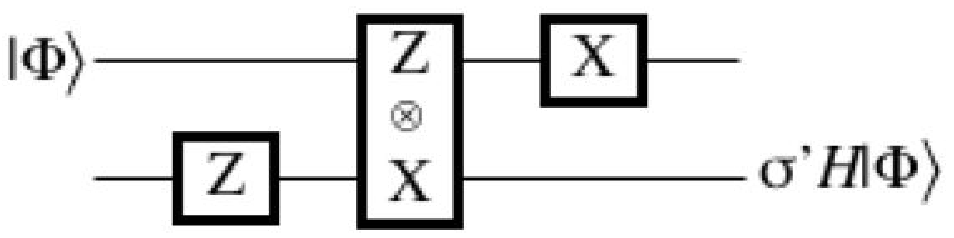}

\emph{\small Figure 4 - Step of simulation of H: $V_1=Id$ and $V_2=H$ \\(note that for all $\sigma$, there exists $\sigma '$ such that $H\sigma=\sigma ' H$)}

$$ $$

\includegraphics[width=0.43\textwidth]{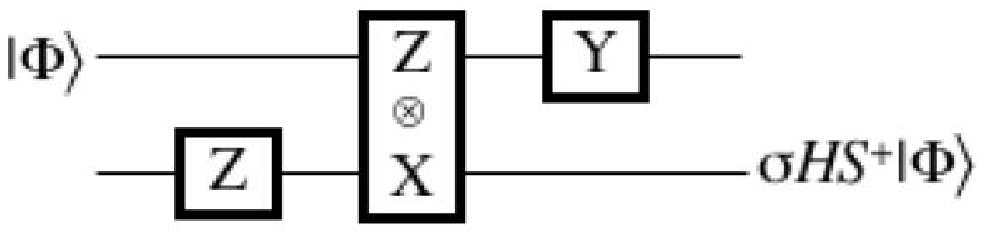}

\emph{\small Figure 5 - Step of simulation of $HS^\dag$: $V_1=S^\dag$ and $V_2=H$.}

$$ $$

\includegraphics[width=0.6\textwidth]{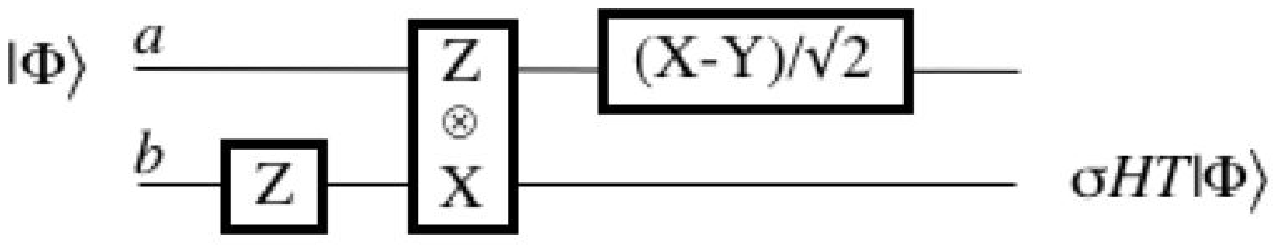}

\emph{\small Figure 6 - Step of simulation of $HT$: $V_1=T$ and $V_2=H$.}

\end{center}

 For a given $2$-qubit register $a,b$ and one auxiliary qubit $c$, the sequence of measurements $\{Z^{(c)},Z^{(a)}\otimes X^{(c)}, Z^{(c)}\otimes X^{(b)}, X^{(c)}\}$ (see fig. 7), simulates the $2$-qubit unitary transformation $CNot$ on the state $\ket{\phi}$ of $a,b$ up to a $2$-qubit Pauli operator which depends on the classical outcomes of the measurements, where:
  
 $$CNot=\left(\begin{array}{cccc}
  1 & 0&0&0\\
  0&1 &0&0 \\
  0& 0&0&1 \\
   0&0 &1&0 \\
\end{array} \right)$$

\begin{center}
\includegraphics[width=0.8\textwidth]{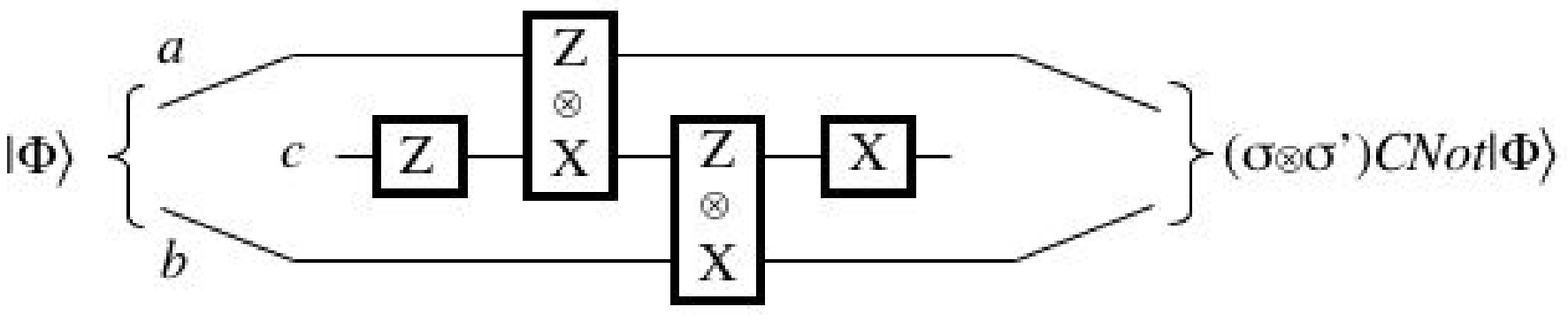}

\emph{\small Figure 7 - Step of simulation of $CNot$}
\end{center}

\section{Universal families of observables}

There exist two types of quantum computation universalities: 
\begin{itemize}

\item A family $\mathcal{F}$ of operators (unitary transformations or measurements) is \emph{quantum universal} iff for all operator $O$, there exists a combination (compositions and tensor products) of some elements of $\mathcal{F}$ which simulates $O$. 
\item A family $\mathcal{F}$ of operators is \emph{approximatively quantum universal} iff for all operator $O_1$ and for all $\epsilon > 0$, there exists an operator $O_2$ and a combination of some elements of $\mathcal{F}$ which simulates $O_2$, with $\vert \vert O_1-O_2\vert \vert \leq \epsilon$. 

\end{itemize}

Since the family of unitary transformations $\mathcal{F}_0=\{H,HT, CNot\}$ is approximatively quantum universal \cite{NC00,KSV}, the family of observables $\mathcal{F}_1=\{Z\otimes X, X, Z, \frac{X-Y}{\sqrt{2}}\}$ is also approximatively quantum universal \cite{P04}. The approximate quantum universality of $\mathcal{F}_1$ is based on the ability to simulate each element of $\mathcal{F}_0$ using elements of $\mathcal{F}_1$ only (see fig. $4$, $6$ and $7$).

\begin{theorem}
 The family of observables $\mathcal{F}_2=\{Z\otimes X, Z,cos(\theta) X +sin(\theta) Y, \theta \in [0\ 2\pi]\}$ is quantum universal.
\end{theorem}

\begin{proof}
Since the family of unitary transformations $\mathcal{F}_3=\{CNot\}\cup\mathcal{U}_1$ (where $\mathcal{U}_1$ is the set of all one-qubit unitary transformations) is quantum universal, the quantum universality of $\mathcal{F}_2$ is reduced to the simulation of each element of $\mathcal{F}_3$.
The proof consists in exhibiting a step of simulation (i.e. a simulation up to a Pauli operator) of each operator of $\mathcal{F}_3$. The stages of correction, omitted in this proof, are presented in \cite{P04}Ê for $\mathcal{F}_1$.

A step of simulation of $CNot $ is presented in figure 7. For a given $U \in \mathcal{U}_1$, $U$ can be decomposed into three successive elementary rotations about the $\hat{z} $, $\hat{x}$ and $\hat{z}$ axes in the Bloch sphere. The elementary rotation $R_{\hat{x}}$ can be expressed using an elementary rotation $R_{\hat{z}}$ and the Hadamard transformation $H$: $$\forall \varphi, R_{\hat{x}}(\varphi)=HR_{\hat{z}}(\varphi)H.$$

Thus for all one-qubit unitary transformation $U$, there exist $\varphi_1,\varphi_2, \varphi_3$ such that $U=R_{\hat{z}}(\varphi_3)HR_{\hat{z}}(\varphi_2)HR_{\hat{z}}(\varphi_1)$, where 

$$R_{\hat{z}}(\varphi)=\left(\begin{array}{cc}
  1 & 0\\
  0 & e^{i\varphi}\\
\end{array} \right)$$

So $U$ can be decomposed into $4$ operators: $$U=\underbrace{(H)}_{U_4}\underbrace{(HR_{\hat{z}}(\varphi_3))}_{U_3}\underbrace{(HR_{\hat{z}}(\varphi_2))}_{U_2}\underbrace{(HR_{\hat{z}}(\varphi_1))}_{U_1}$$
\end{proof}

\noindent A step of simulation of $U_4=H$ is presented in figure $4$, and the simulation of $U_i=HR_{\hat{z}}(\varphi_i), i \in [1,3]$ is obtained using the generalized state transfer of figure $3$. Since $R_{\hat{z}}(\varphi)^\dag Z R_{\hat{z}}(\varphi)=Z$ and $R_{\hat{z}}(\varphi)^\dag X R_{\hat{z}}(\varphi)=cos(\varphi)X-sin(\varphi)Y$, it comes the following step of simulation of $HR_{\hat{z}}(\varphi)$:
\begin{center}
\includegraphics[width=0.65\textwidth]{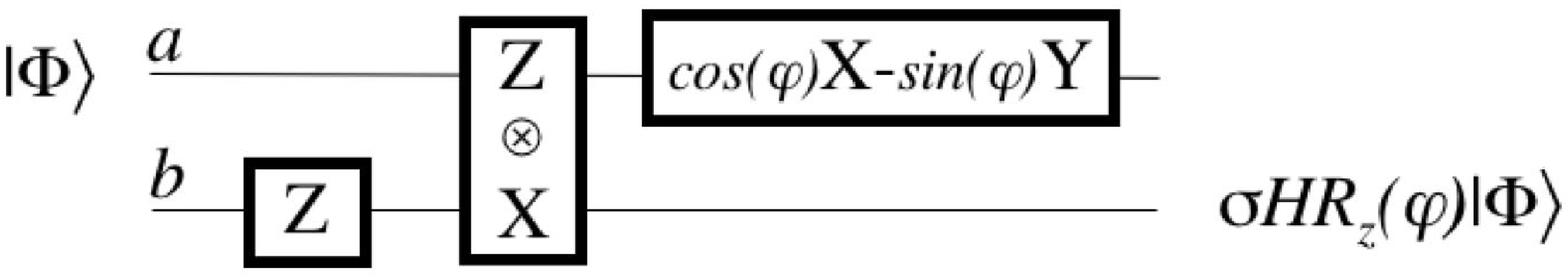}

\emph{\small Figure 8 - Step of simulation of $HR_{\hat{z}}(\varphi)$: $V_1=R_{\hat{z}}(\varphi)$ and $V_2=H$.\\(note that for all $\sigma$, there exists $\sigma '$ such that $H\sigma=\sigma ' H$)}
\end{center}

Thus the simulation of any one-qubit unitary transformation can be decomposed into four unitary transformations such that each of these unitary transformations can be simulated using only observables of $\mathcal{F}_2$. Therefore the family of observables $\mathcal{F}_2=\{Z\otimes X, Z,cos(\theta) X +sin(\theta) Y, \theta \in [0,2\pi]\}$ is quantum universal. 

$\hfill \Box$ 

\section{The secret of the One-Way Quantum Computer is hidden in the initial cluster state}

One-way quantum computation consists in measuring qubit after qubit a lattice of qubits, initially prepared in an entangled state: the \emph{cluster state}. This is a \emph{one-way} computation because the entanglement is consumed step by step. Therefore the creation of the intial cluster state is a crutial point.

In order to create the initial cluster state on a given lattice of qubits, the following preparation is performed:

\begin{itemize}
\item Some qubits of the lattice are input qubits, i.e. qubits which are in an unknown state $\ket{\phi}$, others are auxiliary qubits. Each auxiliary qubit is initialized in the state $\ket{+}=\frac{1}{\sqrt{2}}(\ket{0}+\ket{1})$. 
\item An Ising transformation is applied on the whole lattice. This Ising transformation is equivalent to the application of the 2-qubit unitary transformation Controlled-$Z$ ($C_Z$) on each pair of neighboring qubits, where:
\end{itemize}

 $$C_Z=\left(\begin{array}{cccc}
  1 & 0&0&0\\
  0&1 &0&0 \\
  0& 0&1&0 \\
   0&0 &0&- 1 \\
\end{array} \right)$$

Even if the previous Ising transformation has interesting properties, this unitary transformation has to be simulated using quantum measurements only, in order to get a relevant comparison between the one-way quantum computer and the model based on state transfers. Since any unitary transformation can be simulated using quantum measurements only, each $C_Z$ which composes the Ising transformation can be simulated with measurements, but this simulation needs an additional auxiliary qubit \cite{P04}. Therefore one may wonder if the transformation which creates the initial cluster state can be simulated without additional auxiliary qubits. 

\subsection{Simulation of $C_Z$ on $\ket{\phi}\otimes\ket{+}$ without auxiliary qubit}
\begin{center}
\includegraphics[width=0.6\textwidth]{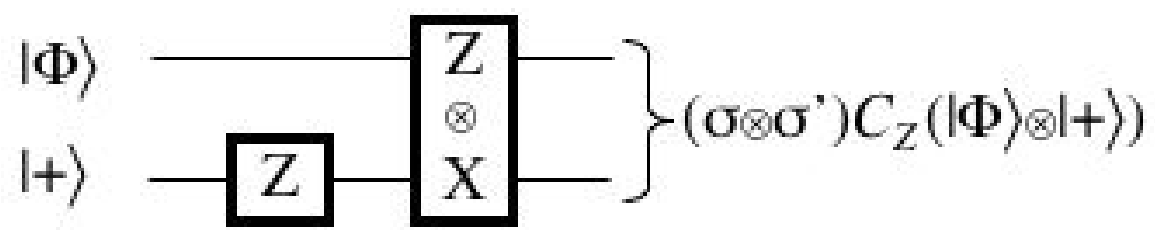}

\emph{\small Figure 9 - Simulation of $C_Z$ on $\ket{\phi}\otimes\ket{+}$ without auxiliary qubit}
\end{center}

\begin{lemma}
For a given qubit $a$ in an unknown state $\ket{\phi}$ and a given qubit $b$ in the state $\ket{+}$, the sequence of measurements $\{Z^{(b)}, Z^{(a)}\otimes X^{(b)}\}$ (see fig. 9) simulates the unitary transformation $C_Z$ on $\ket{\phi}\otimes\ket{+}$ up to a two-qubit Pauli operator.
\end{lemma}

\begin{proof}

If $\ket{\phi}=\alpha\ket{0}+\beta\ket{1}$ and if the outcome of $Z^{(b)}$ is $i\in\{-1,1\}$, then the state $\ket{\psi_1}$ of the register $a,b$ after this measurement is:

$\ket{\psi_1}=( I \otimes \sigma_x^\frac{1-i}{2})[\alpha\ket{00}+\beta\ket{10}]$.

If the outcome of $Z^{(a)}\otimes X^{(b)}$ is $j\in\{-1,1\}$, then the state $\ket{\psi_2}$ of the register $a,b$ after this measurement is:

$\ket{\psi_2}=\frac{1}{\sqrt{2}}( \sigma_z^\frac{1-i}{2}\otimes \sigma_z^\frac{1-j}{2})[\alpha\ket{00}+\alpha\ket{01}+\beta\ket{10}-\beta\ket{11}]$.

Since $C_Z(\ket{\phi}\otimes \ket{+})=\alpha\ket{00}+\alpha\ket{01}+\beta\ket{10}-\beta\ket{11}$, the $2$-qubit unitary transformation $C_Z$ is simulated on $\ket{\phi}\otimes \ket{+}$ up to a $2$-qubit Pauli operator.$\hfill \Box$
\end{proof}

\subsection{Creation of a one-dimensional Cluster State}

In order to create the initial cluster state on a one-dimensional $n$-qubit lattice composed of a unique input qubit, a \emph{cascade} of $C_Z$ can be performed as follow:

\begin{center}
\includegraphics[width=0.43\textwidth]{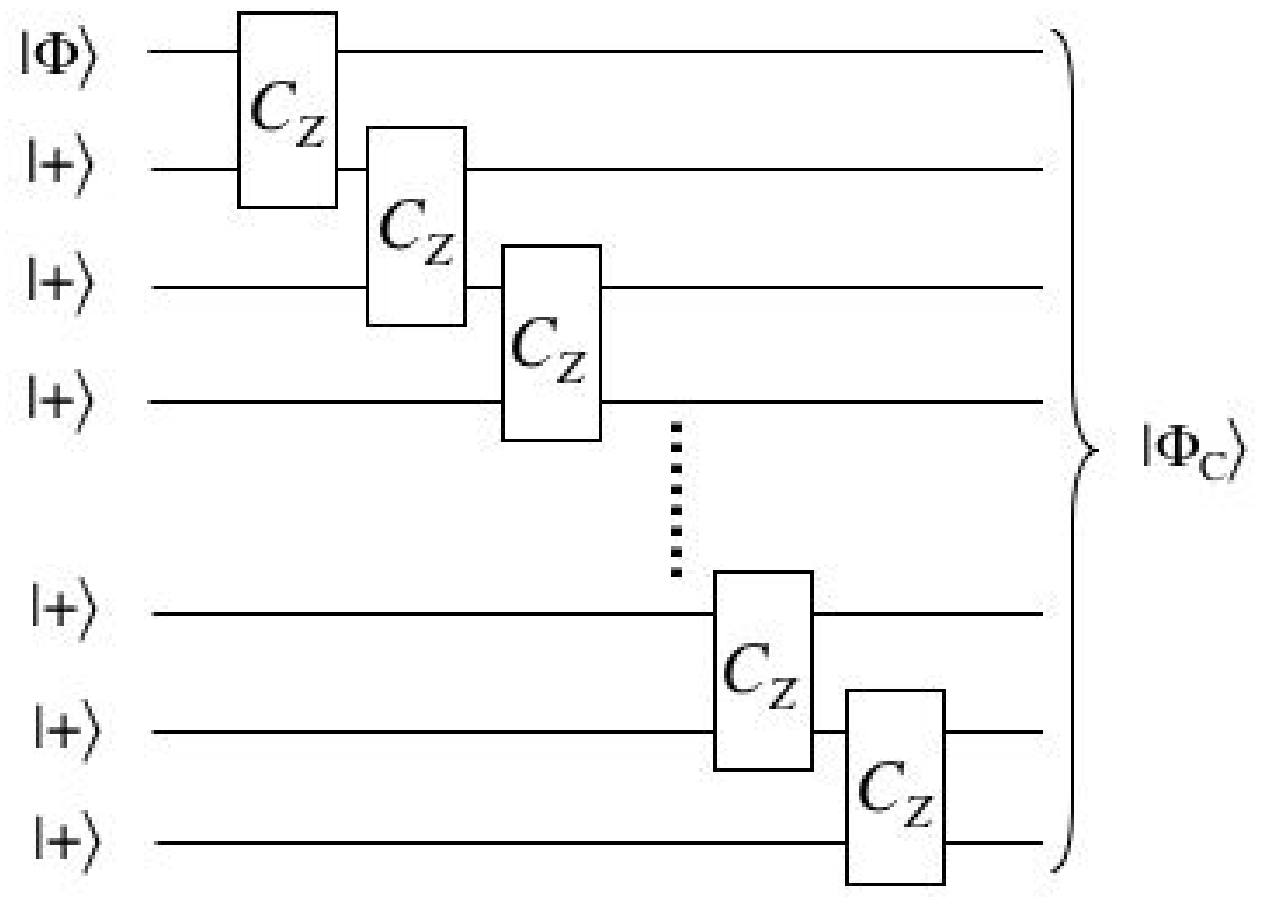}

\emph{\small Figure 10 - Cascade of $C_Z$ for creating the initial cluster state $\ket{\phi_C}$}
\end{center}

For each $C_Z$ of the previous cascade the state of the second input qubit is $\ket{+}$, thus, according to  \emph{Lemma 1}, the previous cascade of $C_Z$ is simulated by the following cascade of measurements. Note that this simulation requires no additional auxiliary qubit. 

\begin{center}
\includegraphics[width=0.47\textwidth]{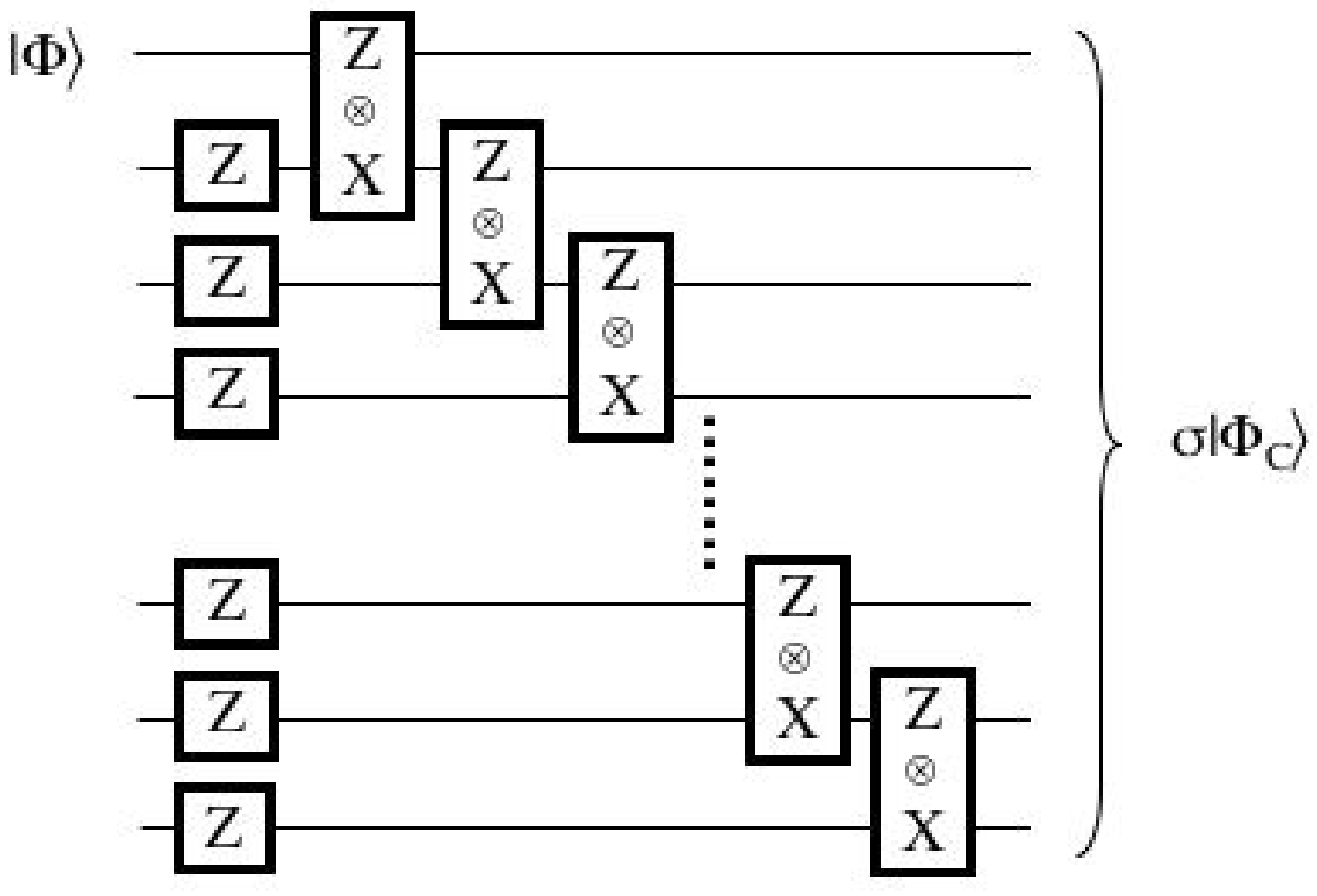}

\emph{\small Figure 11 - Cascade of measurements for creating the initial cluster state $\ket{\phi_C}$ up to a Pauli operator}
\end{center}

\section{Executions on a One-Way Quantum Computer}

An execution on a one-way quantum computer is a sequence of one-qubit measurements on a cluster state. For instance, for a given five-qubit cluster state, if the first qubit is considered as an input qubit $\ket{\phi}$ and if the sequence of measurements $\{X,Y,Y,Y\}$ is performed on the first four qubits (see fig 12), then the state of the last qubit is $\sigma H\ket{\phi}$. Thus the one-way quantum computer of figure 12 simulates the Hadamard transformation. 

\begin{center}
\includegraphics[width=0.25\textwidth]{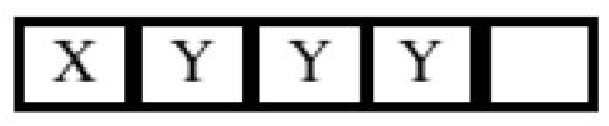}

\emph{\small Figure 12 - Simulation of the Hadamard transformation}
\end{center}

If the phase of preparation of the cluster state and the phase of execution are both represented (see fig $13a$), then the measurements implied in the preparation of the cluster state and those implied in the execution can be decomposed into a succession of generalized state transfers (see fig $13b$). 

\begin{center}
\includegraphics[width=0.37\textwidth]{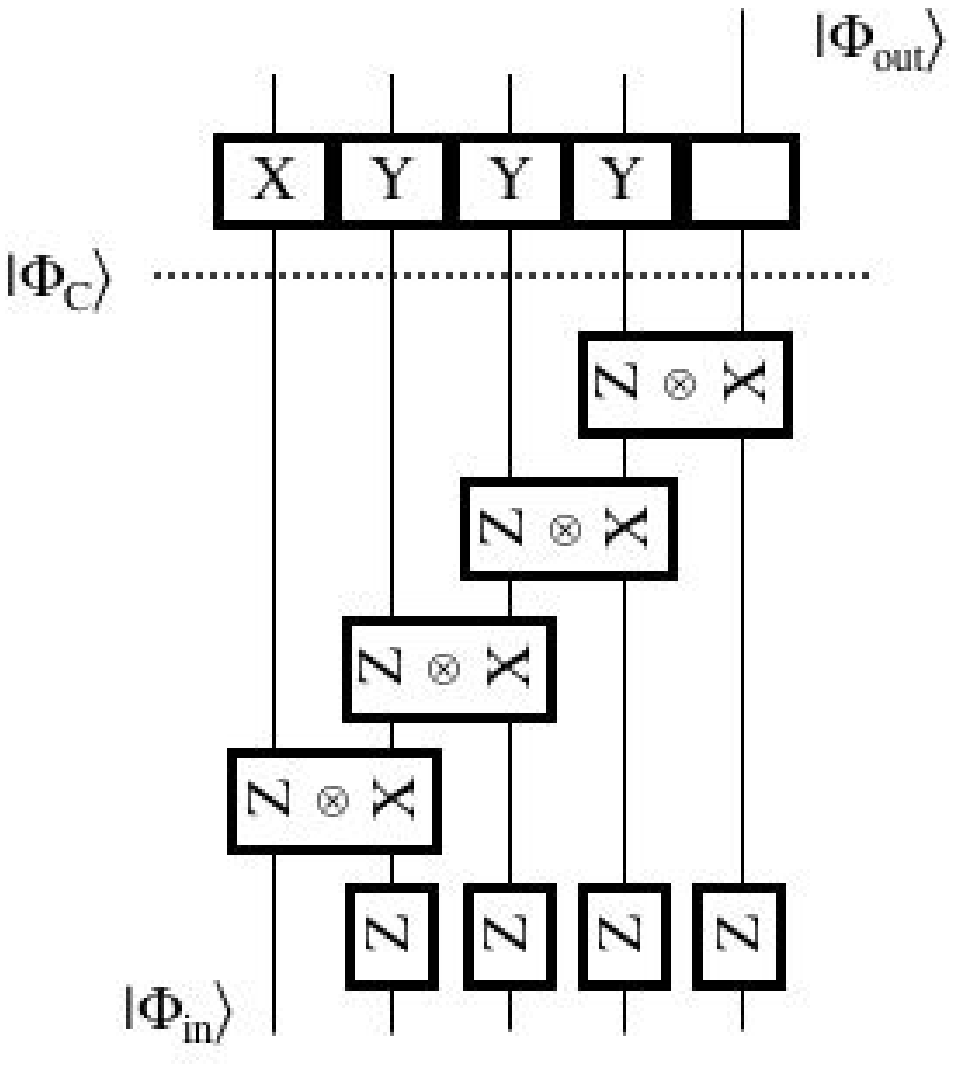}
\includegraphics[width=0.37\textwidth]{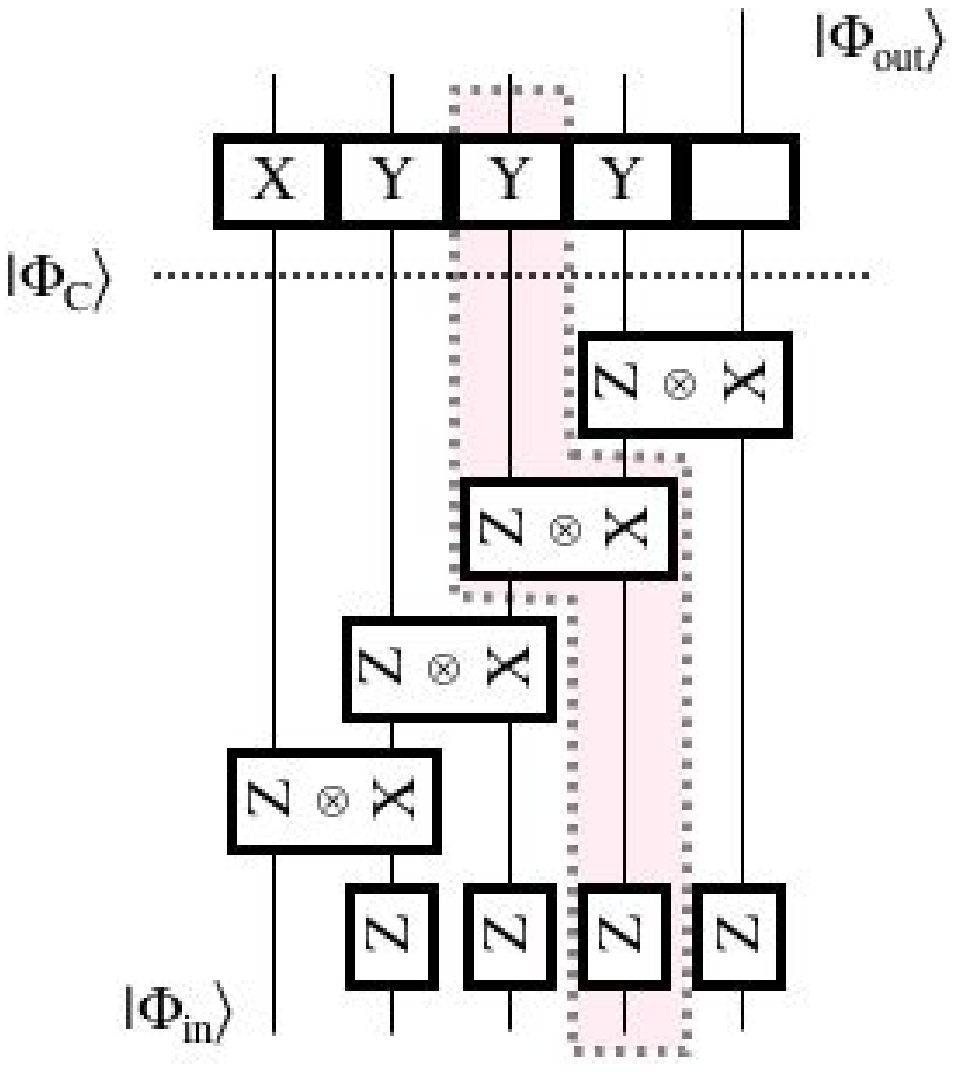}
\includegraphics[width=0.24\textwidth]{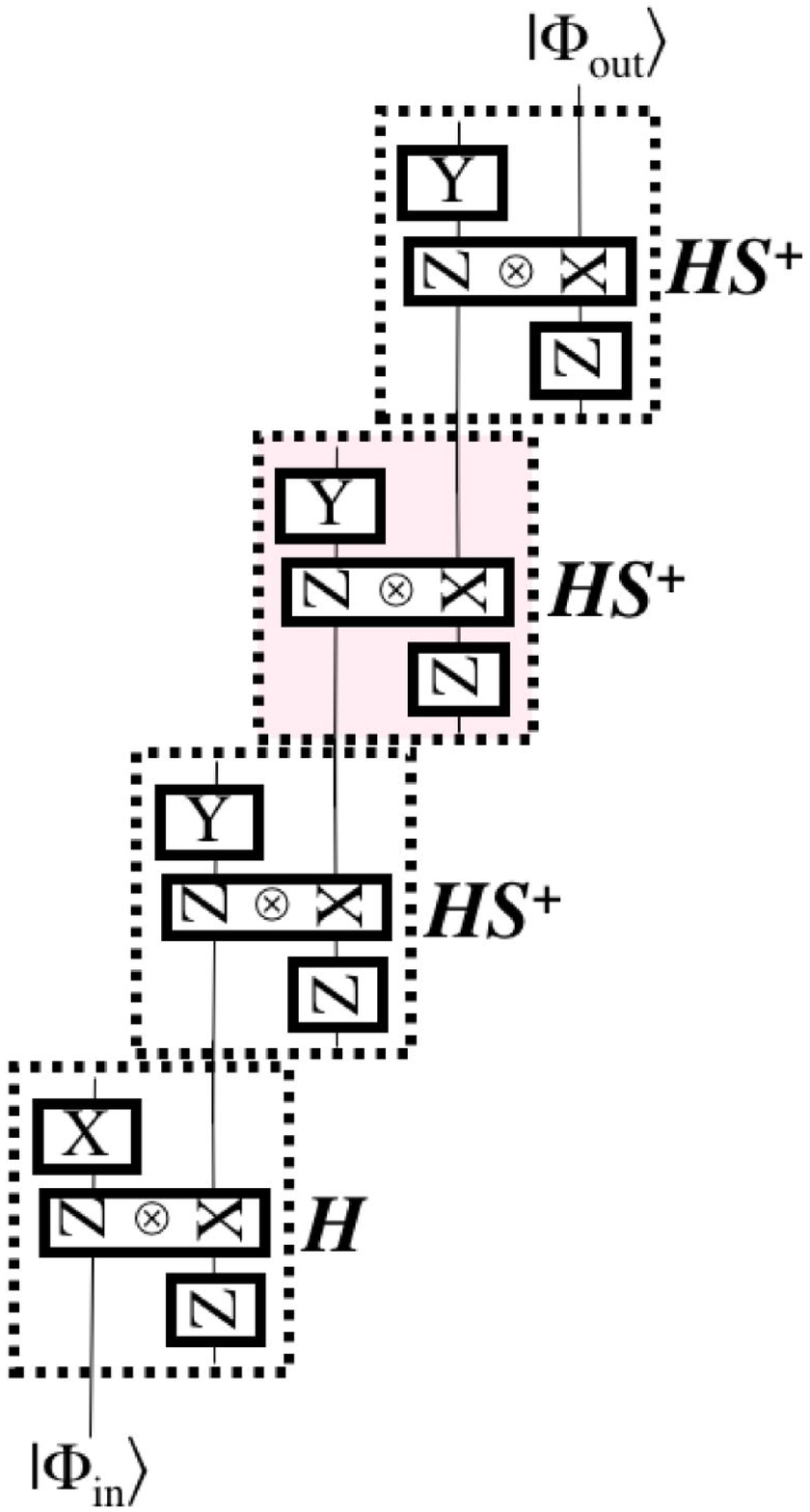}

\emph{\ \ \ \ \ \ \ \ \ (a)\ \ \ \ \ \ \ \ \ \ \ \ \ \ \ \ \ \ \ \ \ \ \ \ \ \ \ \ \ \ \ \ \ (b) \ \ \ \ \ \ \ \ \ \ \ \ \ \ \ \ \ \ \ \ \ \ (c)\ \ \ \ }

\emph{\small Figure 13 - Execution on a one-way quantum computer}
\end{center}

This decomposition offers a natural translation from any one-dimensional one-way quantum computer to quantum computation via projective measurements only. Moreover a straightforward interpretation of the action of any one-dimensional one-way quantum computer is obtained. For instance, the one-way quantum computer of figure 12 can be decomposed (see fig $13c$) into a step of simulation of $H$ (fig 4), and three steps of simulation of $HS^\dag$ (fig 5), thus the action $U$ of this one-way quantum computer is $U=(HS^\dag )(HS^\dag )(HS^\dag )(H)=H$.

\begin{center}
\includegraphics[width=0.25\textwidth]{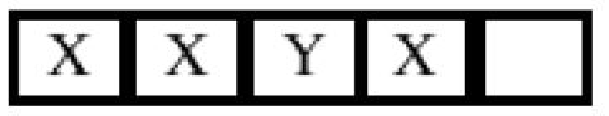}

\emph{\small Figure 14 - Simulation of $S^\dag$}
\end{center}

Similarly, the action $U$ of the one-way quantum computer presented in figure 14 is $U=(H)(HS^\dag )(H)(H)=S^\dag$. 

More generally, the measurements allowed in a one-way quantum computation are in the basis $\mathcal{B}(\varphi)=\{\frac{\ket{0}+e^{i\varphi}\ket{1}}{\sqrt{2}},\frac{\ket{0}-e^{i\varphi}\ket{1}}{\sqrt{2}}\}$ for all $\varphi$. The observable associated with $\mathcal{B}(\varphi)$ is $\mathcal{O}(\varphi)=cos(\varphi)X+sin(\varphi)Y$. Each $\mathcal{O}(\varphi)$-measurement is associated with a generalized state transfer with $V_1=R_{\hat{z}}(-\varphi)$ and $V_2=H$ (see fig 8).

Thus the action $U$ of the one-way quantum computer of figure 15 is $U=HR_{\hat{z}}(-\zeta)HR_{\hat{z}}(-\eta)HR_{\hat{z}}(-\xi)H$.

\begin{center}
\includegraphics[width=0.25\textwidth]{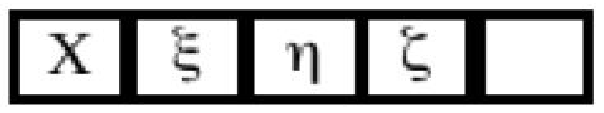}

\emph{\small Figure 15 - Simulation of a general one-qubit unitary transformation}
\end{center}

The connexions between the one-way quantum computer and genera\-lized state transfer may also be used for designing new one-way quantum computers. For instance the one-way quantum computers introduced in figure 16 simulate respectively $H$ and $S$ while requiring less qubits than those introduced by Briegel and Raussendorf \cite{R00,R03}.

\begin{center}
\includegraphics[width=0.11\textwidth]{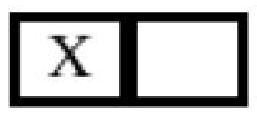}\ \ \ \  \ \ \ \  \ \ \ \  \ \ \ \  
\includegraphics[width=0.20\textwidth]{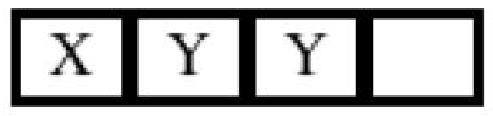}

\emph{\small Figure 16 - Left: simulation of $H$ - Right: simulation of $(HS^\dag )(H S^\dag )(H)=S$}
\end{center}

\section{Conclusion}

In this paper, we have introduced an exact quantum universal family of observables including only one two-qubit observable, while others are one-qubit observables. 

Moreover, the connections established between models of measurement-based quantum computation, permit a natural translation of each one-dimensional one-way quantum computer into a sequence of state transfers. Therefore, by these close connections, the measurement-based quantum computer is unified: a one-way quantum computer is nothing but a quantum computer based on state transfers, in which a large part of the measurements (those independent on the program we want to perform) are grouped in a stage of \emph{initialization}. Note that this initialization can be performed using an Ising transformation, which is unitary. The initialization produces the \emph{cluster state}, on which the rest of the measurements (composed of one-qubit measurements only) are performed in order to complete the computation.

\emph{Final remark. This paper deals with unifying models of quantum computation via measurements only, and with minimizing universal families of observables. It has been submitted to a conference on April $1^{st}$, $2004$. The authors have recently noticed a report posted on arXiv.org, by P. Ali\-feris and D. W. Leung \cite{AL04}, dealing with unifying models of quantum computation via measurements only. The relations among both approaches are certainly worth investigating further.}


\begin{thebibliography}{99}
\bibitem{KSV}A. Y. Kitaev, A. H. Shen and M. N. Vyalyi. {\it Classical and Quantum Computation}, American Mathematical Society, 2002. 
\bibitem{L01} D. W. Leung. {\it Two-qubit projective measurements are universal for quantum computation}, arXiv.org report quant-ph/0111077, 2001.
\bibitem{L03} D. W. Leung. {\it Quantum computation by measurements}, arXiv.org report quant-ph/0310189, 2003.
\bibitem{N01} M. A. Nielsen. {\it Universal quantum computation using only projective measurement, quantum memory, and preparation of the 0 state}, arXiv.org report quant-ph/0108020, 2001.
\bibitem{NC00} M. A. Nielsen and I. L. Chuang. {\it Quantum Computation and Quantum Information}, Cambridge University Press, 2000.

\bibitem{P04} S. Perdrix {\it State Transfer instead of Teleportation in Measurement-based Quantum  Computation
}, arXiv.org report quant-ph/0402204, 2004.

\bibitem{R00} R. Raussendorf and H. J. Briegel. {\it Quantum computing via measurements only
} Phys. Rev. Lett. 86 5188, 2000.
\bibitem{R03} R. Raussendorf, D. E. Browne and H. J. Briegel. {\it Measurement-based quantum computation with cluster states}, arXiv, quant-ph/0301052, 2003.

\bibitem{VC03} F. Verstraete, J.I. Cirac. {\it Valence Bond Solids for Quantum Computation}, arXiv.org report quant-ph/0311130, 2003.

\bibitem{AL04} P. Aliferis, D. W. Leung. {\it Computation by measurements: a unifying picture}, arXiv.org report quant-ph/0404082, 2004.

\end{thebibliography}
\end{document}